\documentclass[]{article}
\usepackage{graphicx}
\usepackage{authblk}
\graphicspath{ {img/} }
\title{Segmentation of points of interest during fetal cardiac assesment in the first trimester from color Doppler ultrasound}
\author[1]{Ruxandra Stoean}
\author[2]{Dominic Iliescu}
\author[1]{Catalin Stoean}
\affil[1]{Romanian Institute of Science and Technology, Cluj-Napoca, Romania}
\affil[2]{Department of Obstetrics and Gynecology, University of Medicine and Pharmacy of Craiova, Craiova, Romania}

\begin{document}

\maketitle

\begin{abstract}
The present paper puts forward an incipient study that uses a traditional segmentation method based on Zernike moments for extracting significant features from frames of fetal echocardiograms from first trimester color Doppler examinations. A distance based approach is then used on the obtained indicators to classify frames of three given categories that should be present in a normal heart condition. The computational tool shows promise in supporting the obstetrician in a rapid recognition of heart views during screening.
\end{abstract}

\section{Introduction}

Fetal cardiac examination is one of the crucial ultrasound assessments during pregnancy towards an early detection of possible congenital heart disease (CHD). CHD occurs in 5 to 9 newborns in 1000 births, of which 18\% die within the first year of life. The disease accounts for 50\% of childhood deaths attributed to congenital malformations \cite{patel}.

The echocardiography can allow immediate intervention after birth, or even intrauterine treatment, and an associated decreased cost of care. In a fetal echocardiography, the physicians look for the absence, distortion or improper positioning of different heart components per se or in relation to one another, as well as for an incorrect flow of blood in the heart. The fetal heart scanning is mainly performed in the second trimester of pregnancy. However, an early detection of the condition should start with the first trimester \cite{hutchinson}, when the pregnancy can be terminated in case of severe abnormalities, and that can be correlated with the findings of the subsequent trimester \cite{jicinska}. While screening for CHD started with the high-risk groups (genetic, environmental factors), it is now generally advocated, due to many cases discovered in low-risk pregnancies \cite{cara}. The prenatal detection rate of CHD by a medical sonographer in the second trimester has been reported between 80-90\% when a standardized protocol is used \cite{iliescu}, \cite{letourneau}. For the first trimester, the guidelines do not impose the obligation of screening for CHD because the heart structures are still underdeveloped, so the detection rate is close to null.

The present paper adopts a traditional segmentation approach based on polygon counting and computation of Zernike moments for first trimester recognition of a normal blood flow in the incipiently formed heart. A distance-based separation into three necessarily present classes of frames is then performed on the base of the extracted features. The preliminary results, while based on a small sample of ultrasound scans, advocate the potential of the computational support for first trimester fetal cardiac assessment.

\section{Problem Description}

The echography approach to the cord necessitates a high degree of expertise of the sonographer. The fetal cardiac sweeping takes into account 5 transversal planes and 3 longitudinal ones. The optimal images are obtained when the fetus has the peak of the cord oriented towards the anterior maternal wall. An unfavorable position necessitates extra time for recording the video or reappointment.

\begin{figure}
	\centering
	\begin{tabular}{cccc}
		\includegraphics[width=0.19\textwidth]{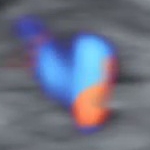} &  \includegraphics[width=0.19\textwidth]{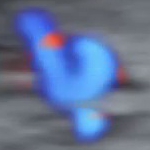} & \includegraphics[width=0.19\textwidth]{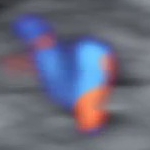} & \includegraphics[width=0.19\textwidth]{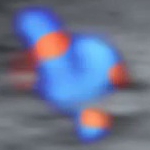}\\
		\includegraphics[width=0.19\textwidth]{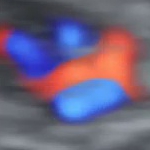} &  \includegraphics[width=0.19\textwidth]{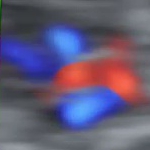} & \includegraphics[width=0.19\textwidth]{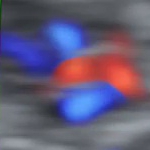} & \includegraphics[width=0.19\textwidth]{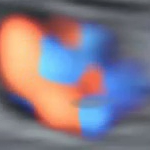}\\	
		\includegraphics[width=0.19\textwidth]{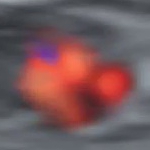} &  \includegraphics[width=0.19\textwidth]{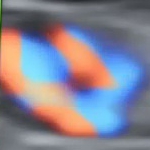} & \includegraphics[width=0.19\textwidth]{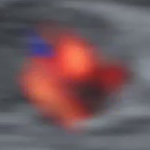} & \includegraphics[width=0.19\textwidth]{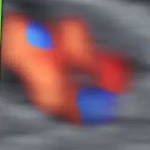}\\
	\end{tabular}
	\caption{Representation for each of the classes of frames that need to be identified. First row contains samples of $V$ in blue, the second illustrates $X$ in the same color and the bottom row shows the $parallel$ red.}
	\label{fig:representations}
\end{figure}

Color Doppler must also accompany the evaluation of the echography views, in order to confirm the normality of the blood flow in the corresponding structures. In red, there is the flow of blood coming towards the transducer and, in blue, the flow of blood going away from the probe. In the first trimester, the fetal heart is barely developed, so only the parallel red flows should be observed, along with the V of the blue flow and sometimes the presence of an X (Figure \ref{fig:representations}).

Accordingly, the goal lies in identifying the "V" and the "X" signatures in the blue representation, as well as finding the two red flows that appear like two parallel lines. In order to treat this object identification task as a classification task, we will also consider another class that corresponds to a different representation than the previously mentioned ones. This will be further referenced as "other" representation. In a usual US video file, most of the frames belong to the $other$ class. On the other extent, the $X$ representations are the ones that appear most seldom and may even be absent.

As it can be observed in the images from Figure~\ref{fig:representations}, the illustrations that belong to the same class are not identical.

\section{Proposed Methodology}

In order to create a machine learning approach, we need to prepare the data for training the model. The format of the data is of US video files. Frames are extracted from each such file and the physicians help in separating the frames into classes. Then the regions of interest are extracted and distinct characteristics are calculated from each frame. A model will be subsequently trained to learn to distinguish between the items from each class. It will be then utilized to recognize the frames that contain regions of interest that are important for the human experts.

Since the labels for the distinct classes are established starting from the shapes and the topology of the identified polygons, we treat the instances for the two colors separately.

\subsection{From Videos to Regions of Interests}

The frames are extracted from each US video file. The training and test sets comprise each frames from different patients for objectivity. The problems are next separated based on the objects of the two distinct colors. For the blue objects, frames from each of the three classes are classified by the physician. Similarly, for the red objects problem, the human expert establishes frames for the parallel lines, as well as for the $other$ class.

\begin{figure}
	\centering
	\begin{tabular}{cc}
		\includegraphics[width=0.29\textwidth]{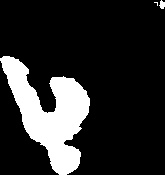} &
		\includegraphics[width=0.69\textwidth]{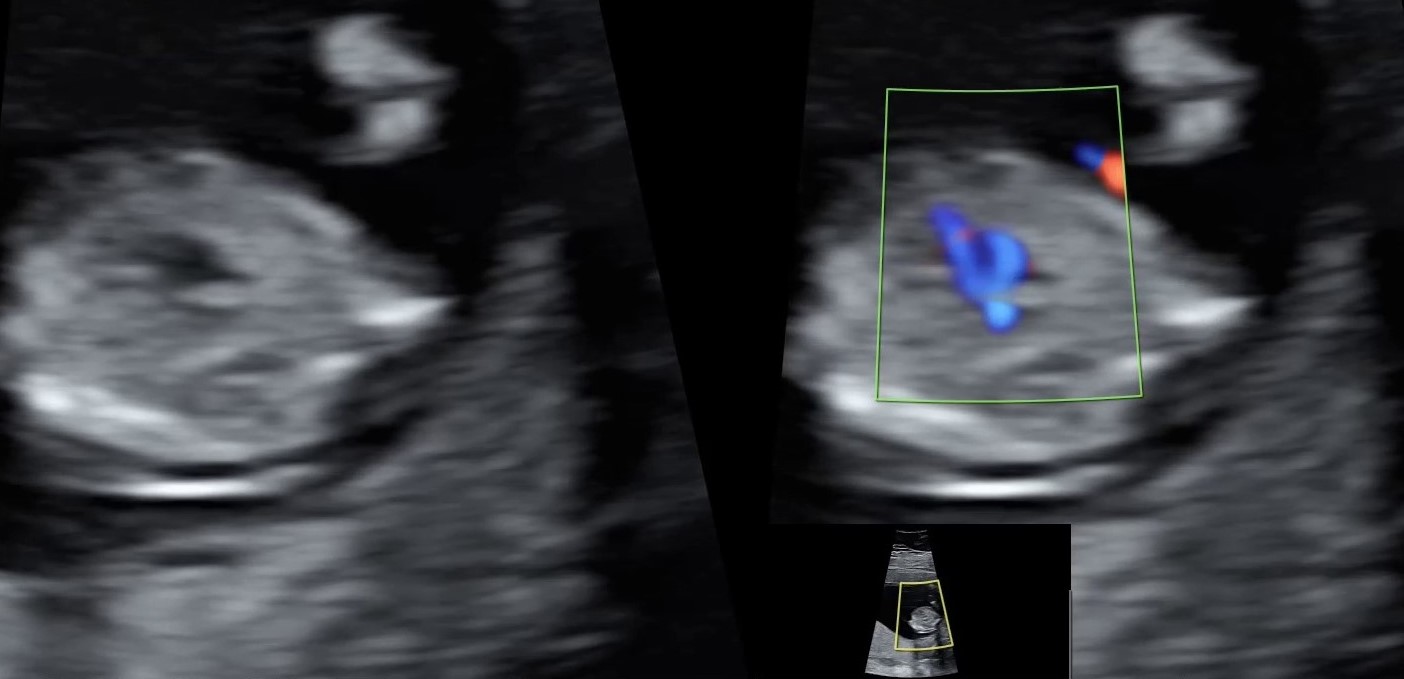}\\	
		\includegraphics[width=0.29\textwidth]{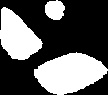} &
		\includegraphics[width=0.69\textwidth]{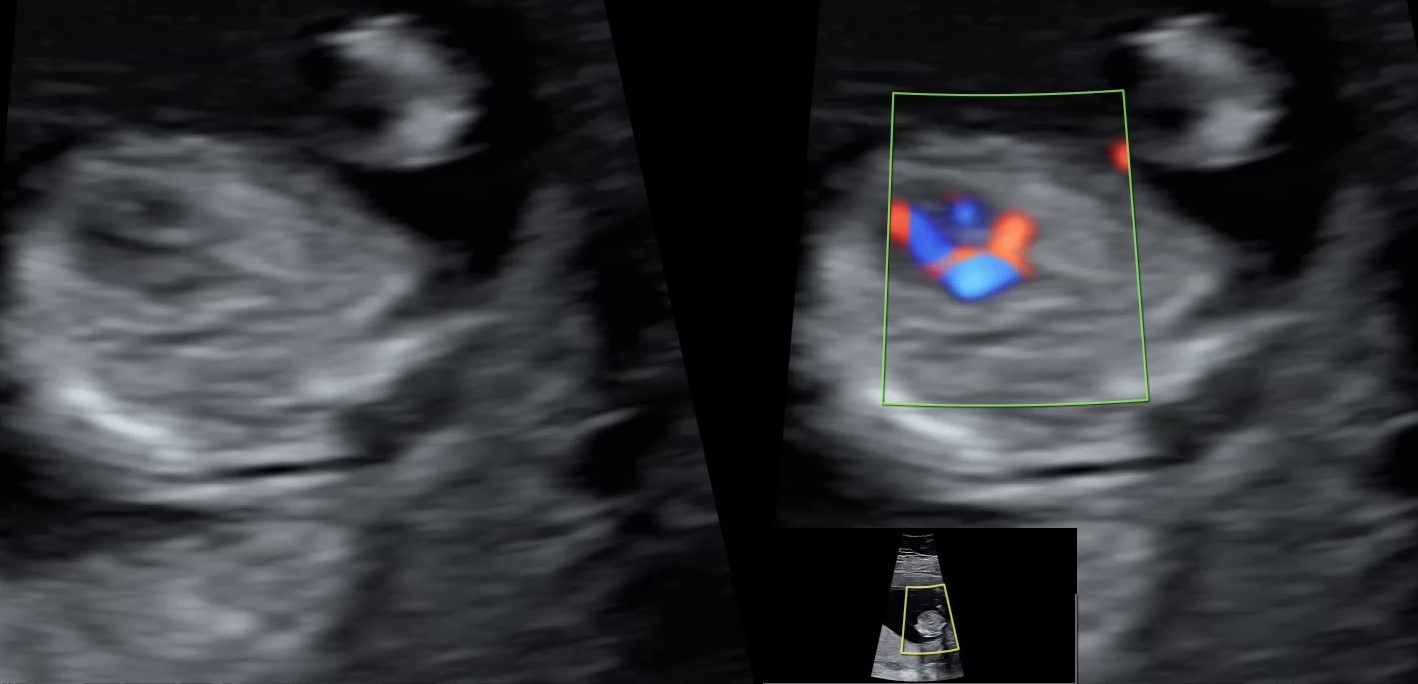}\\
		\includegraphics[width=0.29\textwidth]{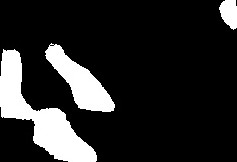} &
		\includegraphics[width=0.69\textwidth]{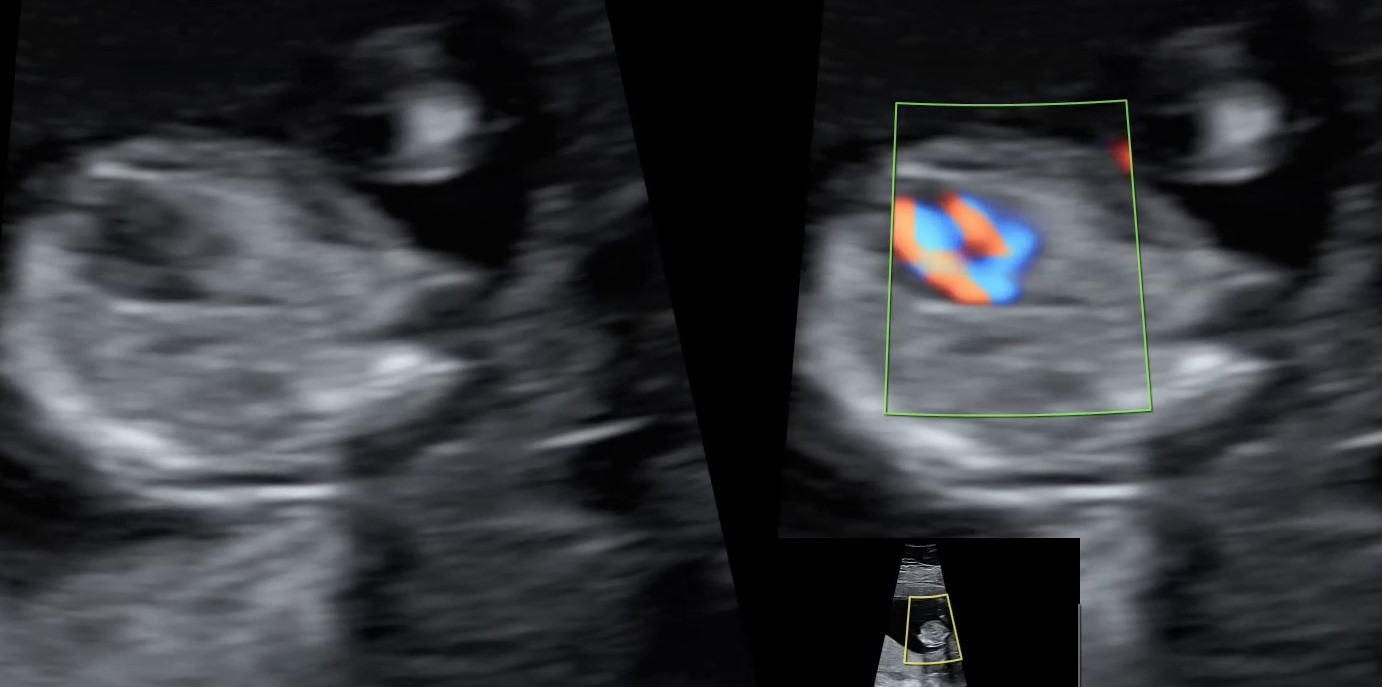}\\
	\end{tabular}
	\caption{Examples of extracting the binary regions of interests (on the left) from the initial images (on the right). From top to bottom, blue $V$, blue $X$ and the $parallel$ red.}
	\label{fig:extractions}
\end{figure}

Binary images are subsequently obtained by cropping from the frames only the part that contains polygons that are either blue (for the blue problem) or red (for the other task). The obtained polygons in the binary images lack nuances that appear in the colored version and the forms that are followed by the human expert no longer appear the same. However, since the same procedure is applied for all the frames, the shapes of the binary polygons become distinctive further in the machine learning procedure.

\subsection{Feature Extraction}

Naturally, not all the cropped binary images have the same sizes. Since the width and height of the images also carry information about the contained shapes, the two characteristics are further saved in the numerical data set that is formed. The number of polygons existing in the image is also imagined to contain information about the class of each image in turn, and is thus further retained. Beside these, there are 25 Zernike moments values that are calculated for each image in turn. Accordingly, each image is described by 28 features in the current approach. Naturally, further characteristics can be added to the numerical data set, like various statistics (e.g. minimum, maximum, average and so one) with respect to the areas, perimeters of the polygons, or even their topology (Delaunay triangles, Voronoi diagrams). Nevertheless, the pre-experimental results of the current approach are already encouraging.

Next, a machine learning approach could be employed for learning the particularities of each class. Still, even using a distance based approach that would assign for each test sample the output of the closest training sample leads to relatively good accuracy. Such a distance-based approach is employed in the current proof of concept study due to the small amount of training data that was available at that moment.

\section{Preliminary Results}

Our preliminary tests refer to a small  data sets of 53 selected slides for training for the $blue$ classification task. There were 20 frames for class $V$, 12 for $X$ and 21 for the $other$ class. The smaller amount for class $X$ is due to the fact that this class can be absent. On the other hand, for $other$ there can be a high number of frames selected, but a balance between the classes for the training part is also desired. These slides were selected in such a manner to cover situations as different as possible, not only with respect to $X$ and $V$, but also among themselves.

All the training slides are selected from video files coming from different patients that those in the test collection. The test samples are not carefully selected to belong to specific classes, but rather the application is made directly on the consecutive slides from the video files. Consequently, the test set contains a large number of slides (that is 230). The physician appreciated that most of the $V$ and $X$ class slides were correctly identified. On the downside, there were also false alarms, since there were some $other$ images that were labeled as $V$. Nevertheless, such an approach would already provide substantial help for a physician, since it would indicate some frames from a US video that would need further attention and would eliminate many other unimportant images that are not helpful, thus reducing the workload and even helping the residents until they gain more experience.

Similar results are obtained for the red problem, where a distribution of 20 samples for each class is used. Again, there were some slides that were mistaken for $parallel$, while their class should have been $other$, but the actual $parallel$ are mostly correctly found.

\section{Conclusions and Future Work}

The current results encourage us to further explore the presented idea. The main concern lies in the preparation of the training, validation and test sets, since the number of frames need to be significantly larger for all classes, in order to have measurable results. Next, on the computational part, the number of descriptive attributes can be enriched with various morphological and topological features. Moreover, a classification model, like a support vector machine, would have more advanced representations that would surely lead to more accurate results. With a substantial increase in the number of available data samples, a deep learning model could be appointed that would be able to perform the feature extraction and classification by its own.

\bibliographystyle{plain}
\bibliography{RSDICS}

\begin{thebibliography}{1}

\bibitem{cara}
Monica Cara, Stefania Tudorache, Roxana Dimieru, Maria Florea, Ciprian Patru,
  and Dominic Iliescu.
\newblock Prenatal first trimester assessment of the heart.
\newblock {\em Annals of Cardiology and Cardiovascular Medicine}, 1(2):1008, 06
  2017.

\bibitem{hutchinson}
Darren Hutchinson, Angela McBrien, Lisa Howley, Yuka Yamamoto, Priya Sekar,
  Tarek Motan, Venu Jain, Winnie Savard, and {Lisa K.} Hornberger.
\newblock First-trimester fetal echocardiography: Identification of cardiac
  structures for screening from 6 to 13 weeks' gestational age.
\newblock {\em Journal of the American Society of Echocardiography}, 2017.

\bibitem{iliescu}
D.~Iliescu, S.~Tudorache, A.~Comanescu, P.~Antsaklis, S.~Cotarcea, L.~Novac,
  N.~Cernea, and A.~Antsaklis.
\newblock Improved detection rate of structural abnormalities in the first
  trimester using an extended examination protocol.
\newblock {\em Ultrasound in Obstetrics \& Gynecology}, 42(3):300--309, 2013.

\bibitem{jicinska}
Hana Jicinska, Pavel Vlasin, Michal Jicinsky, Ilga Grochova, Viktor Tomek,
  Julia Volaufova, Jan Skovranek, and Jan Marek.
\newblock Does first-trimester screening modify the natural history of
  congenital heart disease?
\newblock {\em Circulation}, 135(11):1045--1055, 2017.

\bibitem{letourneau}
Karen~M Letourneau, David~Hughes Horne, Reeni~N Soni, Keith~R. McDonald,
  Fern~C. Karlicki, and Randall~R. Fransoo.
\newblock Advancing prenatal detection of congenital heart disease: A novel
  screening protocol improves early diagnosis of complex congenital heart
  disease.
\newblock {\em Journal of ultrasound in medicine : official journal of the
  American Institute of Ultrasound in Medicine}, 37(5):1073--1079, 2018.

\bibitem{patel}
Neel Patel, Evan Narasimhan, and Anne Kennedy.
\newblock Fetal cardiac us: Techniques and normal anatomy correlated with adult
  ct and mr imaging.
\newblock {\em RadioGraphics}, 37:160126, 06 2017.

\end{thebibliography}

\end{document}